\input epsf.sty
\newdimen\psfigsize
\def\psfigure#1 #2 #3 #4 #5{
    \begin{figure}[tbh]
      \vbox{
        \null\vskip-0.1in\hskip#2
        \epsfxsize=#1
        \epsfbox{#4}
        \vskip -0.1in
        \caption {#5 \label{#3}}
        \vskip 0.0truein plus0.1truein
      }
    \end{figure}
}

\def\nabstar#1{\nabla\kern-0.5pt\smash{\raise 4.5pt\hbox{$\ast$}}
               \kern-4.5pt_{#1}}

\def\drvstar#1{\partial\kern-0.5pt\smash{\raise 4.5pt\hbox{$\ast$}}
               \kern-5.0pt_{#1}}

\def\newline{\relax\ifhmode\null\hfil\break\else\nonhmodeerr@\newline\fi}
\def\frac#1#2{{#1\over#2}}
\def\text#1{{\hbox{\rm #1}}}

\newcommand{\beq}{\begin{equation}}
\newcommand{\eeq}{\end{equation}}
\newcommand{\bea}{\begin{eqnarray}}
\newcommand{\eea}{\end{eqnarray}}
\def\Id{ \mbox{1\hspace{-1.2mm}I} }
\def\BE{\begin{equation}}
\def\EE{\end{equation}}
\def\BA{\begin{eqnarray}}
\def\EA{\end{eqnarray}}
\def\BAN{\begin{eqnarray*}}
\def\EAN{\end{eqnarray*}}

\def\nn{\nonumber\\}

\def\tr{\mathrm{tr}}

\def\gm5{\gamma_5}

\def\det{\mathrm{det}}

\def\BE{\begin{equation}}
\def\EE{\end{equation}}
\def\BA{\begin{eqnarray}}
\def\EA{\end{eqnarray}}
\def\BAN{\begin{eqnarray*}}
\def\EAN{\end{eqnarray*}}

\def\nn{\nonumber\\}

\def\tr{\mbox{tr}}

\def\det{\mbox{det}}

\def\text#1{{\rm #1}}
\documentstyle[twoside,fleqn,espcrc2]{article}
\begin{document}
\title{The Index and Axial Anomaly of a Lattice Dirac Operator}
\author{Ting-Wai Chiu%
\address{Department of Physics, National Taiwan University,
Taipei, Taiwan 106, Republic of China.}
\thanks{%
This work was supported by the National Science Council, R.O.C. under the
grant numbers NSC89-2112-M002-079 and NSC90-2112-M002-021}}

\begin{abstract}

A remarkable feature of a lattice Dirac operator is discussed.
Unlike the Dirac operator for massless fermions in the continuum, 
this Ginsparg-Wilson lattice Dirac operator does not possess topological
zero modes for any topologically-nontrivial background gauge fields,
even though it is exponentially-local, doublers-free, and reproduces
correct axial anomaly for topologically-trivial gauge configurations.

\end{abstract}

\maketitle

\section{INTRODUCTION}

In the continuum, the Dirac operator of massless fermions in a smooth
background gauge field with non-zero topological charge $ Q $ has zero
eigenvalues and the corresponding eigenfunctions are chiral.
The Atiyah-Singer index theorem \cite{Atiyah:1968mp}
asserts that the difference of the number of left-handed and right-handed
zero modes is equal to the topological charge of the gauge field
configuration :
\bea
n_{-} - n_{+} = Q \ .
\label{eq:AS_thm}
\eea
However, if one attempts to use the lattice
to regularize the theory nonperturbatively, then {\it not} every
lattice Dirac operator might
possess topological zero modes
\footnote{So far, it has been confirmed that
overlap Dirac operator \cite{Neuberger:1998fp,Narayanan:1995gw} and its
generalization \cite{Fujikawa:2000my}
can possess topological zero modes with index satisfying (\ref{eq:AS_thm}),
on a finite lattice.}
with index satisfying
(\ref{eq:AS_thm}), even if it is exponentially-local, doublers-free, and
reproduces correct axial anomaly for topologically-trivial gauge backgrounds.
As a consequence, a topologically-trivial lattice Dirac operator
might not realize 't Hooft's solution to the $ U(1) $ problem in QCD, nor
other quantities pertaining to the nontrivial gauge sectors. Nevertheless,
from a theoretical viewpoint, it is interesting to realize that one may
have the option to turn off the topological zero modes of a
lattice Dirac operator, without affecting its correct behaviors
( axial anomaly, fermion propagator, etc. ) in the topologically-trivial
gauge sector. In this talk, I discuss an example of such
lattice Dirac operators \cite{Chiu:2001bg}
and show that it does not possess topological zero modes
for any topologically-nontrivial gauge configurations
satisfying a very mild condition, Eq. (\ref{eq:nonzero}).

\section{A LATTICE DIRAC OPERATOR}

Consider the lattice Dirac operator \cite{Chiu:2001bg}
\bea
\label{eq:DcD}
D = a^{-1} D_c ( \Id + r D_c )^{-1} \ , \hspace{4mm} r = \frac{1}{2c} \ ,
\eea
with
\bea
\label{eq:Dc}
D_c &=& \sum_{\mu} \gamma^\mu T^{\mu} \ ,
\hspace{2mm} T^\mu = f t^\mu f \ ,
\eea
\bea
\label{eq:f}
f = \left( \frac{2c}{\sqrt{t^2 + w^2} + w } \right)^{1/2}, \
t^2 = -\sum_{\mu} t^\mu t^\mu \ .
\eea
Here $ \gamma^\mu t^\mu $ is the naive lattice fermion operator
and $ -w $ is the Wilson term with a negative mass $ -c $ ( $ 0 < c < 2 $ )
\bea
\label{eq:tmu}
t^\mu (x,y) &=& \frac{1}{2} [ U_{\mu}(x) \delta_{x+\hat\mu,y}
                       - U_{\mu}^{\dagger}(y) \delta_{x-\hat\mu,y} ] \ , \\
\label{eq:Umu}
U_\mu(x) &=& \exp
       \left[ i a g A_\mu \left( x+\frac{a}{2}\hat\mu \right) \right] \ , \\
\label{eq:w}
w(x,y) &=& c - \frac{1}{2} \sum_\mu \ [ \ 2 \delta_{x,y} +  \nn
       & &     - U_{\mu}(x) \delta_{x+\hat\mu,y}
               - U_{\mu}^{\dagger}(y) \delta_{x-\hat\mu,y} \ ] \ ,
\eea
where the Dirac, color and flavor indices have been suppressed.
Note that the $ D_c $ defined in Eq. (\ref{eq:Dc})
can be regarded as a symmetrized version of that constructed in
Ref. \cite{Chiu:1999hz}, for vector gauge theories.

In the free fermion limit, (\ref{eq:DcD}) gives
\bea
\label{eq:Dp}
D(p) = D_0(p) + i \sum_{\mu} \gamma_\mu D_\mu(p) \ ,
\eea
where
\bea
\label{eq:D0p}
D_0(p) &=& \frac{c}{a} \left(1-\frac{w(p)}{\sqrt{t^2(p)+w^2(p)}} \right) \ , \\
\label{eq:Dmu}
D_\mu(p) &=& \frac{c}{a} \frac{\sin( p_\mu a )}{\sqrt{t^2(p)+w^2(p)}} \ , \\
\label{eq:t2}
t^2(p) &=& \sum_{\mu} \sin^2( p_\mu a ) \ , \\
\label{eq:wp}
w(p)   &=& c -  \sum_{\mu} [ 1 - \cos( p_\mu a ) ] \ .
\eea

Evidently, both $ D_0(p) $ and $ D_{\mu}(p) $ are analytic functions for
all $ p $ in the Brillouin zone.
Thus
\bea
\label{eq:Dx}
D(x) = \int \frac{d^4 p}{(2\pi)^4} e^{ i p \cdot x } D(p) \
\eea
is exponentially-local in the position space.
The exponential locality of $ D $ in the free fermion limit immediately
suggests that $ D $ is also exponentially-local for sufficiently
smooth background gauge fields.

It is easy to check that $ D(p) $ is doublers-free and
has the correct continuum behavior, i.e.,
in the limit $ a \to 0 $,
\bea
\label{eq:Dp0}
D(p) \sim i  \sum_{\mu} \gamma_\mu p_\mu + O( a p^2 )  \ .
\eea

Further, $ D $ is $\gamma_5$-hermitian,
\bea
\label{eq:g5_hermit}
D^{\dagger} = \gm5 D \gm5 \ ,
\eea
and it breaks the chiral symmetry according to the
Ginsparg-Wilson relation \cite{Ginsparg:1982bj}
\bea
\label{eq:gwr}
D \gamma_5 + \gamma_5 D = 2 r a D \gamma_5 D \ .
\eea
Thus $ D $ satisfies the necessary requirements for a decent
lattice Dirac operator.

The GW relation (\ref{eq:gwr}) immediately implies
each zero mode of $ D $ has a definite chirality, and the index of $ D $
is equal to the sum of the axial anomaly $ \tr[ a \gm5 D(x,x) ] $
over all sites \cite{Hasenfratz:1998ri},
\bea
\label{eq:index_rel}
\mbox{index}(D) = n_{-} - n_{+} = r \sum_x \tr[ \gm5 a D(x,x) ] \ ,
\eea
where the trace "tr" runs over the Dirac, color and flavor space.

However, the index relation (\ref{eq:index_rel}) does {\it not}
necessarily imply that $ D $ can possess topological zero modes
with the index satisfying (\ref{eq:AS_thm}).
In fact, the GW Dirac operator (\ref{eq:DcD}) always has
\bea
\label{eq:top_trivial}
n_{+} = n_{-} = \sum_{x} \tr[ a \gm5 D(x,x) ] = 0 \ ,
\eea
for any topologically-nontrivial gauge background, even though $ D $
is exponentially-local, doublers-free, $\gamma_5$-hermitian, and
has correct continuum behavior. The proof is as follows.

\section{A PROOF OF THE ABSENCE OF TOPOLOGICAL ZERO MODES}

From (\ref{eq:g5_hermit}) and (\ref{eq:gwr}), we have
\bea
\label{eq:normal}
D^{\dagger} + D = 2 r a D^{\dagger} D = 2 r a D D^{\dagger} \ .
\eea
Thus $ D $ is normal and $\gm5$-hermitian. Then the eigenvalues
of $ D $ are either real or in complex conjugate pairs.
Each real eigenmode has a definite chirality,
but each complex eigenmode has zero chirality.
Further, the sum of the chirality of all real
eigenmodes is zero ( chirality sum rule ) \cite{Chiu:1998bh}.
Now the eigenvalues of $ D $ (\ref{eq:DcD})
fall on a circle in the complex plane, with center $ ( c/a, 0 ) $
on the real axis, and radius of length $ c/a $. Then the
chirality sum rule reads
\bea
\label{eq:chi_sum_rule}
n_{+} - n_{-} + N_{+} - N_{-} = 0 \ ,
\eea
where $ n_{+} ( n_{-} ) $ denotes the number of zero modes of
positive ( negative ) chirality, and $ N_{+} ( N_{-} ) $
the number of nonzero ( eigenvalue $ 2c/a $ ) real eigenmodes
of positive ( negative ) chirality.

The chirality sum rule (\ref{eq:chi_sum_rule}) asserts that each
topological zero mode must be accompanied by a nonzero real eigenmode
with opposite chirality, and vice versa. ( Note that both topological zero
modes and their corresponding nonzero real eigenmodes are {\it robust}
under local fluctuations of the gauge background, thus one can easily
distinguish them from those trivial zero and nonzero real eigenmodes which
are unstable under local fluctuations of the background ).

It follows that if $ D $ cannot have any nonzero real eigenmodes in
topologically nontrivial gauge backgrounds, then $ D $
cannot possess any topological zero modes.

From (\ref{eq:DcD}), any zero mode of $ D $
is also a zero mode of $ D_c $, and vice versa.
However, a nonzero real ( eigenvalue $ 2c/a $ ) eigenmode of $ D $
corresponds to a pole ( singularity ) in the spectrum of $ D_c $, since
$ D_c = D ( \Id - r a D )^{-1} $, the inverse tranform of (\ref{eq:DcD}).

Therefore, if the spectrum of $ D_c $ does {\it not} contain any
poles ( singularities ) for a topologically-nontrivial gauge background,
then $ D $ {\it cannot} have any nonzero real eigenmodes,
thus {\it no} topological zero modes.

Now we consider topologically-nontrivial gauge configurations
satisfying the condition\footnote{It should be emphasized that we have
{\it not} found any {\it robust} nontrivial gauge configuration violating
(\ref{eq:nonzero}), on a finite lattice. Thus, it is likely that
the measure of the nontrivial gauge configurations {\it not} satisfying
(\ref{eq:nonzero}) is {\it zero}.}
\bea
\label{eq:nonzero}
\det ( \sqrt{t^2 + w^2 } + w ) \ne 0 \ .
\eea
Then $ f $ exists, and $ D_c $
(\ref{eq:Dc}) is well-defined ( without any poles ). It follows that
$ D $ (\ref{eq:DcD}) cannot have topological zero modes
for any topologically-nontrivial gauge configurations
satisfying (\ref{eq:nonzero}). This completes the proof.

\section{AXIAL ANOMALY}

From (\ref{eq:top_trivial}), the topological triviality of
$ D $ (\ref{eq:DcD}) implies that it
cannot reproduce correct axial anomaly for topologically-nontrivial
backgrounds. Nevertheless, since $ D $ is exponentially-local, doublers-free
and has correct continuum behavior, these conditions are sufficient to ensure
that it reproduces continuum axial anomaly for topologically-trivial
gauge backgrounds. The axial anomaly of (\ref{eq:DcD}) has been
calculated \cite{Chiu:2001ja} in weak coupling perturbation theory,
up to $ O(g^4) $ of the gauge coupling $ g $,
for toplogically-trivial gauge configurations. It has been shown that
the axial anomaly recovers the topological charge density in the
continuum limit, i.e.,
\BAN
{\cal A} (x) &=& \tr[ \gamma_5 ( \Id - r a D ) (x,x) ] \nn
&=& \frac{g^2}{32\pi^2} \sum_{\mu\nu\lambda\sigma}
 \epsilon_{\mu\nu\lambda\sigma} \tr( F_{\mu\nu} F_{\lambda\sigma} ) + O(a) \ ,
\EAN
where
$ F_{\mu\nu} = \partial_\mu A_\nu - \partial_\nu A_\mu + i g [A_\mu, A_\nu] $.

\section{CONCLUSIONS }

For some years, it has been taken for granted that if a Ginsparg-Wilson
lattice Dirac operator has correct axial anomaly for the trivial gauge
sector, then it must also reproduce continuum axial anomaly for the
nontrivial sectors. However, the lattice Dirac operator (\ref{eq:DcD})
provides a counterexample, and suggests that this common conception may
{\it not} be justified.

In general, given a topologically-proper
lattice Dirac operator, it can be
transformed into a topologically-trivial lattice Dirac operator which
is identical to the topologically-proper one in the free fermion
limit. On the other hand, given a topologically-trivial GW
Dirac operator, it remains an
interesting question how to transform it into a topologically-proper one.

If one insists that the topologically zero modes of a lattice Dirac operator
are crucial for lattice QCD to reproduce the low energy hadron phenomenology,
then one should assure that a Ginsparg-Wilson lattice Dirac operator is
indeed {\it topologically-proper}, before it could be employed for lattice
QCD computations.
However, there might be a very slight possibility that lattice QCD with
{\it topologically-trivial} quarks could reproduce low energy hadron
phenomenology. These issues deserve further studies.

\end{document}